\documentclass{llncs}

\usepackage[T1]{fontenc}
\usepackage[utf8]{inputenc}
\usepackage{todonotes}
\usepackage{url}
\usepackage{setspace} 
\usepackage{graphicx}
\usepackage{multirow}

\begin{document}
	
	\title{A Complete Year of User Retrieval Sessions in a Social Sciences Academic Search Engine} 
    
	\titlerunning{A Complete Year of User Retrieval Sessions}  
	%
	\author{Philipp Mayr, Ameni Kacem}
	\authorrunning{Mayr and Kacem} 
	\institute{GESIS -- Leibniz Institute for the Social Sciences, Cologne, Germany,\\
		\email{[philipp.mayr,ameni.sahraoui]@gesis.org}
	}
	
	\maketitle              

	\begin{abstract}		
In this paper, we present an open data set extracted from the transaction log of the social sciences academic search engine sowiport. The data set includes a filtered set of 484,449 retrieval sessions which have been carried out by sowiport users in the period from April 2014 to April 2015. We propose a description of interactions performed by the academic search engine users that can be used in different applications such as result ranking improvement, user modeling, query reformulation analysis, search pattern recognition. 

     \end{abstract}
	
    \keywords{Whole Session Retrieval, Information Behavior, Session Log \\ Analysis, User Session Data, Social Sciences Users}

\section{Introduction} 
   
Every Digital Library (DL) system generates huge amounts of usage data and DL operators often face the problem of not being able to report about the real usage on an expressive level that is moreover understandable for laymen. Reporting average statistics like number of unique sessions, page impressions, amount of actions and even click-through rates is not enough because these numbers cannot represent and explain the underlying pattern of the information behavior of DL users.
Exploratory search in DLs and academic search engines \cite{Carevic2017} is a rewarding research environment for interactive IR researchers because evolving searches with complex search tasks can be observed much easier compared to web search where searchers often jump into different websites. In DLs, users typically stay in the system and work with the variety of facilities it offers. This is due to the fact that state-of-the-art DLs offer dozens of possibilities to navigate and interact with the search system \cite{Hienert2015,Fuhr2007}. 
Our motivation in proposing this data set is grounded in the observation that in the field very few open data sets which support whole session investigation exist. To the best of our knowledge there is no open data set available from academic search engines or DLs with full coverage of whole session information. Among the available data sets, we find the most famous evaluation campaign TREC (Text REtrieval Conference) which proposed TREC Session\footnote{http://trec.nist.gov/data/session.html} \cite{Kacem17} and Interactive\footnote{http://trec.nist.gov/data/interactive.html} tracks. In fact, one way to enhance the development and evaluation of information-seeking systems is to propose shareable data sets in order to facilitate the collaboration within an interdisciplinary team including developers, computer scientists, and behavioral experts who work together in order to explore new ideas and propose improvements \cite{KellyDP09}.

Consequently, with the proposed data set we want to support DL developers and IR researchers to work on the analysis of whole retrieval sessions. These practitioners need such data sets to propose methods and techniques which allow us to examine search steps, analyze usage data, understand the underlying information behavior covered in search sessions that are performed by geographically distributed persons.

\section{Related Work}  

Interactive information retrieval (IIR) refers to a research discipline that studies the interaction between the user and the search system. In fact, researchers have moved from considering only the current query to consider the user's past interactions. Research approaches aim to understand the user search behavior in order to improve the ranking of results after submitting a query and enhance the user experience with an IR system. Thus, they study concepts such as search strategies \cite{Carevic:2016,Carevic2017}, search term suggestions \cite{Hienert:2016}, communities' detection \cite{Akbar:2012}, personalization of search results, recommendation's impact \cite{Hienert:2016}, user’s information needs frequency and change.
Many interactive IR models have been proposed in the literature (e.g. \cite{Ellis:1989}) that describe the user’s behavior by different steps (stages) of information seeking and interacting with an information retrieval system. 
In order to evaluate and analyze such models and approaches log analysis has been introduced. In \cite{Peters:1993}, the authors proposed a detailed overview of the history and development of transaction log analysis by examining possible applications and features analysis. Jones et al. \cite{Jones:1998} investigated transaction logs for the Computer Science Technical Reports Collection of the New Zealand DL. The authors analyzed query complexity, query terms change, sessions frequency and length.    
	
\section{Dataset}
\label{sec:dataset}
Sowiport\footnote{http://www.sowiport.de} is a DL for the Social Sciences that contains more than nine million records, full texts and research projects included from twenty-two different databases whose content is in English and German \cite{Hienert2015}.
This data set \textbf{Sowiport User Search Sessions Data Set  (SUSS)}\footnote{To download the dataset: http://dx.doi.org/10.7802/1380} \cite{Mayr2016} contains individual search sessions extracted from the transaction log of sowiport. The data was collected over a period of one year (between 2nd April 2014 and 2nd April 2015). The web server log files and specific JavaScript-based logging techniques were, first, used to capture the user behavior within the system. Then, the log was heavily filtered to exclude transactions performed by robots and short interactions limited to one action per session. After that, all transaction activities are mapped to a list of 58 different user actions which cover all types of activities and pages that can be carried out/visited within the system (e.g. typing a query, visiting a document, selecting a facet, exporting a document, etc.). For each action, a session id, the date stamp and additional information (e.g. query terms, document ids, and result lists) are stored. Based on the session id and date stamp, the step in which an action is conducted and the length of the action is included in the data set as well. The session id is assigned via browser cookies and allows tracking user behavior over multiple search sessions. Session boundaries were specified after a threshold period indicating a period of inactivity and thus the end of the session. In our data set this threshold is equal to 20 minutes. Thus, in the data set we find 484,449 individual search sessions and a total of 7,982,427 log entries.

\section{Preliminary analysis} 

In this section, we present, first, a descriptive analysis of the SUSS data set regarding sessions, users, and searches. These analyses are not following concrete research questions but are intended to show the richness of this open data set.  

\subsection{Description of Actions} 

Searching sowiport can be performed through an \textit{All fields} search box (default search without specification), or through specifying one or more field(s): title, person, institution, number, keyword or year.
The users' main actions are described in Table~\ref{tab:actions}. We grouped the main actions into two categories: "Query"-related and "Document"-related actions. Another categorization of actions was proposed in \cite{Hienert:2016} by specifying search interactions and successive positive actions. 

\begin{table}
\centering
\caption{Main actions performed by users in sowiport} 
\label{tab:actions}
\begin{tabular}{|l|l|p{7cm}|r|}
\hline
Category & Action & Description & Frequency\\
\hline
 \multirow{8}{*}{Query} & query\_form & Formulating a query & 179,964 \\
    & search & A search result list for any kind of search &  848,556\\
    & search\_advanced  &  A search with the advanced settings that can limit the search fields, information type, etc. & 103,432 \\
    & search\_keyword & A search for a keyword & 43,608 \\
    & search\_thesaurus & Usage of the thesaurus system & 71,599 \\
    & search\_institution & A search for an institution & 13,104 \\
    & search\_person & A search for a specific person (author/editor) & 93,083\\
\hline 
\multirow{8}{*} {Document} & view\_record & Displaying a record in the result list after clicking on it & 1,344,361\\
    & view\_citation & View the document's citation(s) & 24,994 \\
    & view\_references & View the document's references & 2,086 \\
    & view\_description & View the document's abstract & 86,752\\
    & export\_bib & Export the document through different formats & 27,229 \\
    & export\_cite & Export the document's citations list & 27,385 \\
    & export\_mail & Send the document via email & 10,987  \\
    & to\_favorites & Save the document to the favorite list & 5,431 \\
\hline
\end{tabular}
\end{table}


\subsection{Users and Sessions}
Given the data set described in Section~\ref{sec:dataset}, we first analyze the user types. A user can perform a search and submit a query to sowiport without signing up. Registered users can keep the search history, add a document to favorites and create favorite lists according to their interests. 
We found 1,509 registered users who performed 3,372 unique sessions (0.69\%). 
The rest of the sessions in sowiport were performed by non-registered users (99.31\%).   



\subsection{Investigation of Actions}
Main user actions as described before can be categorized into actions regarding either search queries or documents. These actions are used in different scales in the data set. Query-related actions represent 29.84\% while document-related actions represent 35.79\% of the total amount of actions. The rest of actions contain navigational interactions such as logging in the system, managing favorites, and accessing the system pages.

\begin{figure}[h!]
\begin{center}
	\includegraphics[scale=0.6]{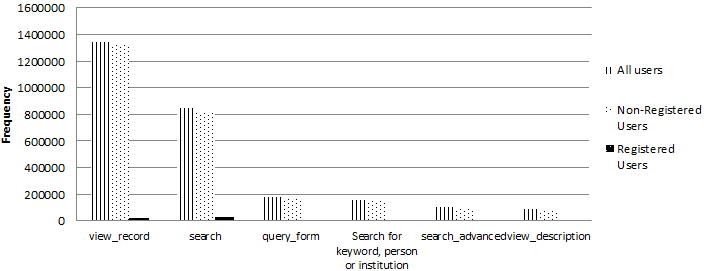}
	\caption{Frequency distribution of the six most performed action groups}
	\label{fig:TopActions}
\end{center}
\end{figure}

Figure~\ref{fig:TopActions} shows the frequencies of the top six most used actions by the users in the data set. We notice that the actions \textit{"view\_record"} and \textit{"search"} are the most used ones before \textit{"query\_form"} and \textit{"search\_{keyword, person, institution}"}. 

In Table~\ref{tab:SessionSample}, we show a specific session, the user's ID and the actions' label and length in seconds. 
In this session, the user with ID \textit{41821} started with logging into the system and then submitted a query describing his/her information need (\textit{query\_form}). After getting the result list, labeled as \textit{resultlistids} and viewing a document, the user performed additional searches (\textit{searchterm\_2}), and displayed some results' content (\textit{view\_record}). Finally, he/she checked the external availability of a result (\textit{goto\_google\_scholar}). We notice that the user spent more than 40\% of the time reading documents' content.

\begin{table}[!htbp]
\centering
\caption{Sample of a session search for a specific user}
\label{tab:SessionSample}
\begin{tabular}{|c|p{3cm}|p{3.5cm}|l|}
\hline
User ID & Date & Action label & Action length (s)\\
\hline
\multirow{9}{*} {41821} & 2014-10-28 16:08:46 & goto\_login   & 1   \\
& 2014-10-28 16:09:13 & query\_form           & 22  \\
& 2014-10-28 16:09:35 & search                & 10  \\
& 2014-10-28 16:09:35 & resultlistids         & 10  \\
 & 2014-10-28 16:09:45 & view\_record          & 31  \\
& 2014-10-28 16:09:45 & docid                 & 31  \\
& 2014-10-28 16:10:16 & view\_record          & 392 \\
& 2014-10-28 16:16:48 & search                & 10  \\
& 2014-10-28 16:16:48 & searchterm\_2         & 10  \\
& 2014-10-28 16:16:48 & resultlistids         & 10  \\
& 2014-10-28 16:16:58 & view\_record          & 9   \\
& 2014-10-28 16:17:07 & goto\_google\_scholar & 0  \\
\hline
\end{tabular}
\end{table}

\begin{figure}[h!]
\begin{center}
	\includegraphics[scale=0.6]{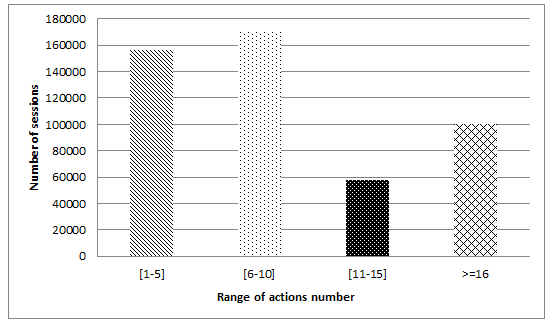}
	\caption{Distribution of the number of actions contained in a session}
	\label{fig:ActionsSessions}
\end{center}
\end{figure}

In Figure~\ref{fig:ActionsSessions}, we display the number of actions per session. We note that the average number of actions per session is 16 and only sessions with a minimum of one action are considered in this data set. We conclude, from this figure, that the number of sessions with less than 16 actions (n=384,087) is much larger than the number of sessions having over 16 actions (n=100,360).

\section{Future work} 
For academia there is a need for open data sets which provide information about the variety of retrieval sessions and help to study and understand the abstract information behavior and common scan paths of academic users in a DL. In fact, session log provision and investigation open opportunities to enhance DLs' systems and to offer new services. Some possible future work based on our proposed data set can be outlined as follows: finding and studying abstract user groups like exhaustive or effective users; modeling academic users; analyzing reformulation and refining strategies; identifying various search phases like starting; chaining, browsing and differentiating; task characterization and prediction; personalization of search results according to the user behavior within search sessions.

\section{Acknowledgement}
This work was funded by Deutsche Forschungsgemeinschaft (DFG), grant no. MA 3964/5-1; the AMUR project at GESIS together with the working group of Norbert Fuhr. 
The AMUR project aims at improving the support of interactive retrieval sessions following two major goals: improving user guidance and system tuning.

\bibliographystyle{splncs}
\bibliography{bibdb.bib}
\end{document}